\documentclass[prl,aps,twocolumn,amssymb,showpacs]{revtex4}
\usepackage{amsmath}
\usepackage{graphicx}
\usepackage{graphics}
\usepackage{epsfig}

\begin{document}

\title{Diffusion of small particles in a solid polymeric medium.}
\author{F. Camboni, A.Koher and I.M. Sokolov}
\affiliation{Institut f\"ur Physik, Humboldt-Universit\"at zu Berlin, Newtonstr. 15, D-12489 Berlin, Germany}

\date{\today}

\begin{abstract}
We analyze diffusion of small particles in a solid polymeric medium 
taking into account a short range particle-polymer interaction. The system is
modeled by a particle diffusion on a ternary lattice where the sites occupied 
by polymer segments are blocked, the ones forming the hull of the chains correspond to
the places at which the interaction takes place, and the rest are voids, in which
the diffusion is free. In the absence of interaction the diffusion coefficient shows only a weak dependence 
on the polymer chain length and its behavior strongly resembles usual site percolation. In presence of
interactions the diffusion coefficient (and especially its temperature dependence) shows a non-trivial
behavior depending on the sign of interaction and on whether the voids and the hulls of the chains
percolate or not. The temperature dependence may be Arrhenius-like or strongly non-Arrhenius, depending on parameters.
The analytical results obtained within the effective medium approximation are in qualitative agreement with those of
Monte Carlo simulations.
\end{abstract}

\maketitle

\section{Introduction}
The literature treating the problem of diffusion of small 
molecules in solid polymeric media is surprisingly limited when compared with
the huge amount of results obtained for diffusion in solids in general. 
This is even more surprising if one takes into account the enormous role polymeric materials 
play as encapsulants and isolating materials in technical devices.
Furthermore, the thermodynamics of polymeric solutions and the dynamics of polymers in 
solutions are well understood since the seminal works by Flory \cite{F} and Huggins \cite{H}.
The situation usually considered is the one in which polymer 
molecules constitute the solute of the solution.  
Only few times the roles have been inverted and polymers have been taken
as solvent molecules forming a matrix in which small solute particles are let to diffuse. 
Early works done in this direction \cite{KMF,Fu} analyze 
the concentration dependence of the small solute diffusivity   
through experimental adsorption and desorption curves. 
In particular, in \cite{Fu}, Fujita concludes with the necessity 
of theoretical and experimental investigations of the ``characteristic differences'' 
between the cases of a good or bad solubility.  
Up to the authors' knowledge, this request has not been satisfied yet. 
Aim of the present work is to give a partial answer by providing 
a qualitative analysis of the way the particle-phobic or 
particle-philic nature of the polymer chains affects the diffusion process.

More specifically, we consider a set of particles diffusing in an amorphous solid polymeric medium, 
in a model being a close relative of a classical Flory-Huggins model of polymer solutions. 
In the present work we adopt the ternary lattice representation corresponding to a polymer-solvent-void system 
close to the one proposed in Ref.\cite{BF}:
In the two variants of the model considered we take a site of a lattice to represent a polymer segment, an interaction site in the vicinity
of a segment, or to be empty. The concentration of solute molecules is considered low, and their interaction with each other is neglected.
  
In the first variant of the model polymers are represented by chains of occupied sites and their nearest neighbors
are considered as interaction sites. Sites not belonging to either of these two categories 
are considered as voids. This lattice model is exactly the one we use in simulations. 
Analytical calculations refer to a simpler mean field Flory-Huggins-like model, 
built by disassembling the chains and letting polymer segments, interaction and empty sites fill the space in a completely 
random fashion at given concentrations. The situations are discussed in depth in Section 2. 
Details of analytical calculations are given in Section 3 with a particular attention 
to the variations to be made with respect to the conventional effective medium technique.  
In Section 4 the interaction between the polymers and the small solute molecules is temporarily switched off and the
model is reduced to a pure percolation problem in the presence of polymer chains. 
This is done in order to estimate the error introduced by the mean field approach and 
the dependence of the diffusion coefficient on the chain length. 
Section 5 is devoted to the role of interaction sites, and Sections 6 contains our conclusions.

\section{The model}

\begin{figure}[h]
\begin{center}
\includegraphics[width=8.0cm]{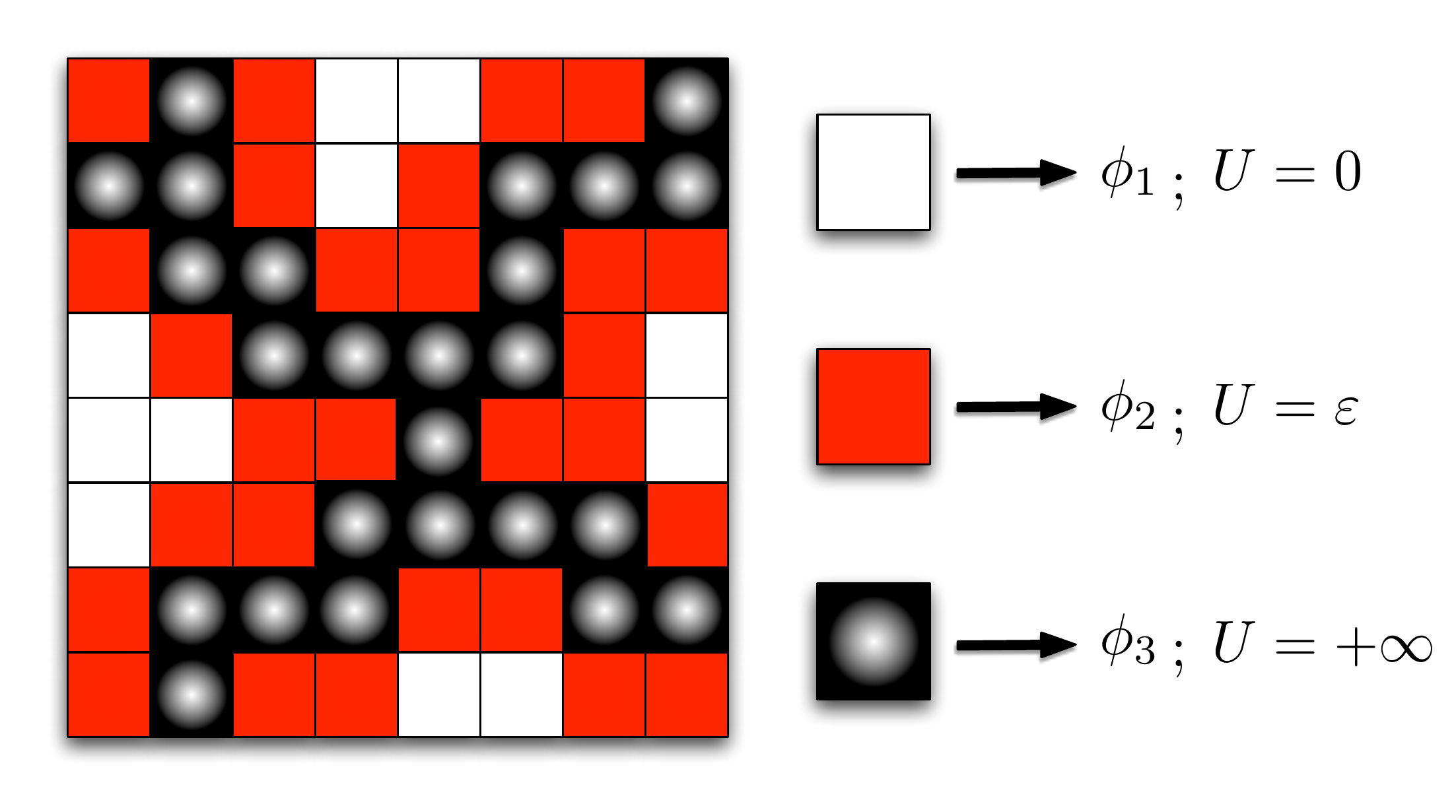} 
\caption{(color online) The ternary lattice of the initial model and different kinds of cells with corresponding concentrations and energy values.}
\label{fig:Real}
\end{center}
\end{figure}

We model our solid polymeric matrix by a three-dimensional cubic lattice 
on which the chains are modeled as phantom random walk
chains of length $l$. This chain conformation corresponds to the 
Gaussian nature of chains in melts from which our solid matrix is
obtained by quenching. The whole matrix is considered as static: no chain motion is taken into account. 
After the system is created, the sites of the lattice occupied 
by chains are considered impenetrable for small solute molecules. 
The whole system is then modeled by a ternary random potential landscape. 
The corresponding lattice is outlined in Figure \ref{fig:Real}. 

The sites occupied by polymer segments are impenetrable for solute molecules 
(hard core interaction, interaction energy $U=\infty$) and represented as black sites in Fig. \ref{fig:Real}. 
The number concentration of these sites is $\phi_3 = M_3/M$ where $M_3$ 
is their total number and $M$ is the volume (total number of sites) of the lattice. 

The particle-polymer interaction is considered to take place
only if the molecule occupies a site which is a nearest neighbor of 
the one occupied by a polymer segment. 
The particle-polymer interaction at these sites corresponds to 
the interaction energy $U=\varepsilon$ whose sign 
fixes the nature of the force experienced by the particles: 
if $\varepsilon$ is negative, this interaction is attractive; 
if $\varepsilon$ is positive, the interaction is repulsive. 
These interaction sites are represented in red in Figure \ref{fig:Real} 
and their number concentration is $\phi_2$.

Remaining sites are considered as simple voids with energy $U=0$ 
and where particles perform a free motion not being subjected to any force. 
The number concentration of these sites is $\phi_1 = 1 - \phi_2 - \phi_3$ and they are represented in white. 
Our system is thus represented by a random (but correlated) ternary lattice with the sites assigned energies $U_i$ which 
take the values $0,\varepsilon$, or $\infty$ for the white, red and black sites respectively. 
In this medium the small molecule diffusion is numerically simulated 
as a nearest-neighbor random walk with transition rates between the
sites given by the corresponding energy differences: 
\begin{equation}\label{TR}
w_{ij}=w_0 e^{-\frac{\beta}{2}(U_i - U_j)}.
\end{equation} 
The constant rate $w_0$ defining the time unit of the process is set to unity in all simulations,
$\beta$ is the usual $1/K_B T$ term 
and $K_B$ is the Boltzmann constant. 

\begin{figure}[h]
\begin{center}
\includegraphics[width=8.0cm]{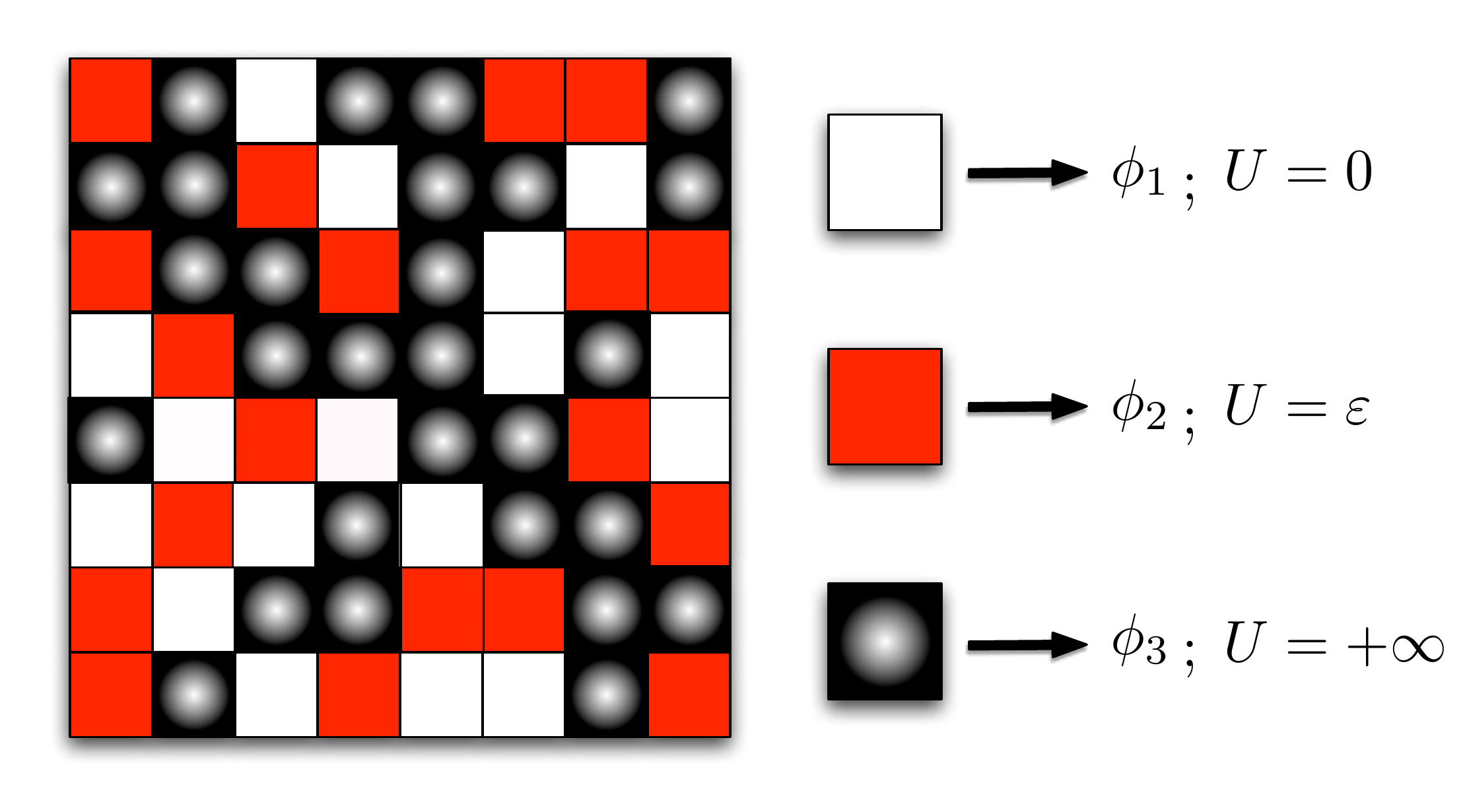} 
\caption{(color online) Mean field ternary lattice and different kinds of cells with corresponding concentrations and energy values.}
\label{fig:MeanField}
\end{center}
\end{figure}

The analytical calculations are performed within a simplified model which strongly
resembles the classical Flory-Huggins model (\cite{F,H,BFM,BF}) 
used for description of thermodynamical properties of polymeric solutions, 
in which the number concentrations of the sites occupied by polymer segments is
kept, but the correlations between their positions (necessarily introduced by the existence of chains) are fully neglected. 
This model corresponds to filling the lattice at random with black, red and white sites at given number concentrations. 
In this way each lattice site 
is assigned an energy value $U_i$ which can take one of the three values $0,\varepsilon$, or $\infty$ 
at random, with probabilities $\phi_k$. 
The existence of an infinite cluster of black sites, which we need to preserve the solidness of the system, 
is guaranteed by taking the concentration $\phi_3$ above the percolation threshold 
which is known to be approximately 0.32 for the three-dimensional simple cubic lattice.
We denote this construction as mean field lattice and represent it in Figure \ref{fig:MeanField}. 
The diffusion on this mean field lattice is then treated using the effective medium approximation for a diffusion in a 
random potential landscape, as discussed in Section 3. The mean field / effective medium results are compared with the results of direct numerical simulations
discussed above, and show qualitatively similar behavior. 

\section{Effective medium approximation for diffusivity}

The particles' motion in a random potential landscape is described via the usual master equation
\begin{equation}
\dot{q}_i = \sum_j \left(w_{ij}q_j - w_{ji} q_i \right),
\label{Meq0}
\end{equation}
where $q_i$ is the probability for a particle to be at a site $i$ at time $t$ 
and $w_{ij}$ is the transition rate from site $j$ to site $i$
given by equation (\ref{TR}) for $j$ and $i$ nearest neighbors and 
equal to zero otherwise.
For the sake of generality calculations will be referring to the $d$-dimensional case.

\begin{figure}[h]
\begin{center}
\includegraphics[width=8.0cm]{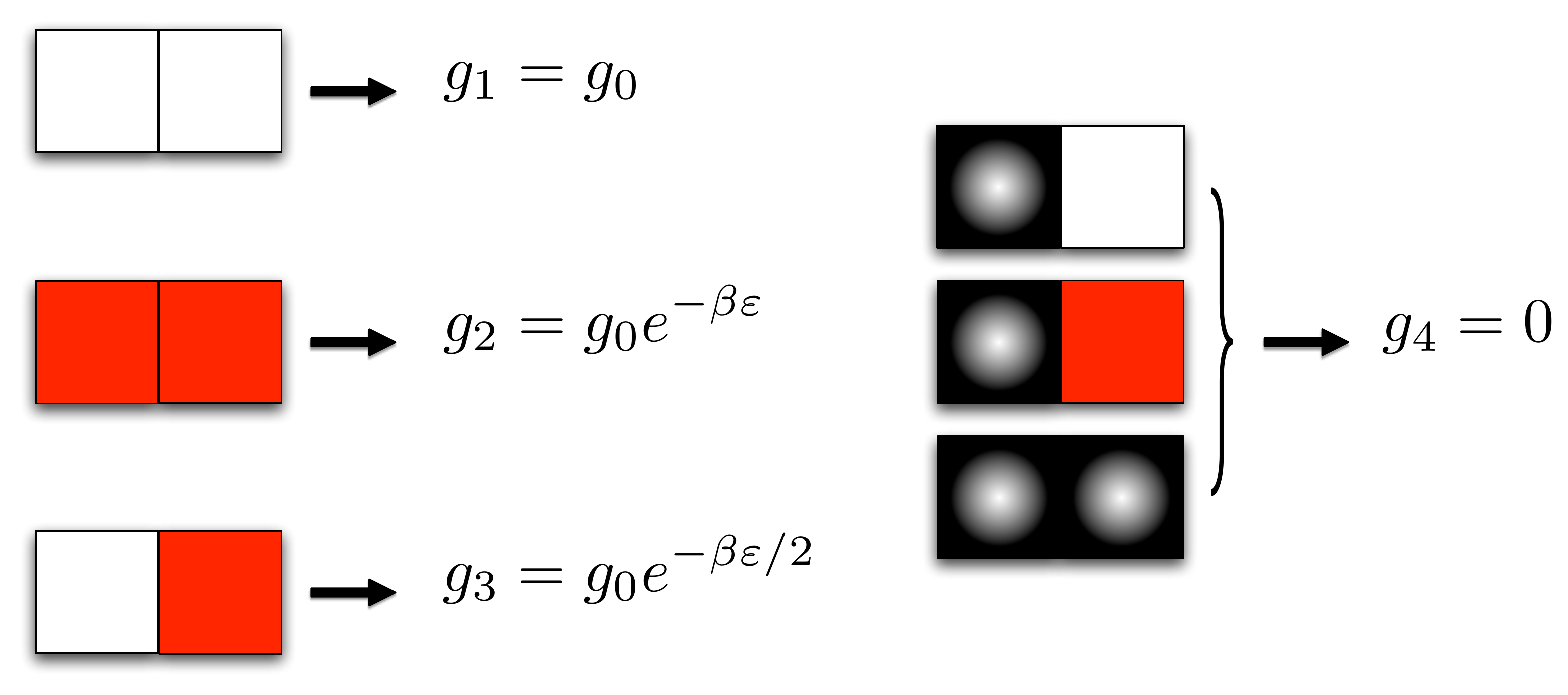}
\caption{(color online) Bond conductivities for the corresponding site couples.}
\label{fig:YLegend}
\end{center}
\end{figure}

We multiply both sides of equation (\ref{Meq0}) by the number 
of particles $N$ and obtain the master equation 
for the site mean number or ``concentration" function
$n_i=Nq_i$.
\begin{equation}
\dot{n}_i = \sum_j \left(w_{ij}n_j - w_{ji}n_i \right).
\label{Meq1}
\end{equation}
Assuming the existence of an equilibrium state,
the transition rates are naturally linked through the detailed balance condition at equilibrium
$w_{ij}n_j^0 = w_{ji}n_i^0$, where $n_i^0=Nq_i^0$ and $q_i^0 \propto \exp{(-\beta U_i)}$ 
is the equilibrium probability to find the
particle at site $i$. Thus one can introduce the symmetrized rates   
$g_{ij}$ being the properties of a bond of a lattice, 
\begin{equation}
g_{ij} = w_{ij}n_j^0 = w_{ji}n_i^0 = g_{ji} = g_0 e^{-\frac{\beta}{2}(U_i + U_j)} 
\label{G}
\end{equation}
with
\begin{equation}
g_0=\frac{N w_0}{Z(\vec{\phi},\varepsilon)} 
\label{g0}
\end{equation}
where $Z(\vec{\phi},\varepsilon)$ is the 
normalization factor of $q_i^0$ (the partition function for the small particles equilibrium distribution)
and $\vec{\phi}$ is the triplet $(\phi_1, \phi_2, \phi_3)$.
Then the analogy between the diffusion and the
electric conduction in a random medium can be used \cite{BG,DS,CS}:
the corresponding diffusion coefficient is connected with the macroscopic conductivity 
$\langle g \rangle_{em}$ of a disordered lattice with bond
conductivities $g_{ij}$ via \cite{CS,D}
\begin{equation}
D_{em} =  a^2 \frac{\langle g \rangle_{em}}{ \left\langle n_i^0\right\rangle} = a^2 \frac{\left\langle w_{ji} \exp(-\beta U_i) \right\rangle_{em}}{ \left\langle \exp(-\beta U_i)\right\rangle}. 
\label{Dem}
\end{equation}
with $a$ the lattice spacing. 
Our system exhibits four different bond conductivity values  
depending on the color of the sites involved. These are
\begin{equation}
g_1 = g_0;  \quad g_2 = g_0 e^{-\beta \varepsilon};  \quad 
g_3 = g_0e^{-\beta \varepsilon/2};  \quad g_4 = 0.
\end{equation}
Figure \ref{fig:YLegend} gives an overall view of this situation.\\

The effective conductivity $\langle g \rangle_{em}$ can then be calculated within the effective medium approximation (EMA).
There is however a subtlety in application of the effective medium approximation to site models like ours.
The genuine continuous EMA of Ref. \cite{B} and its lattice variant Ref. \cite{K} describe well the behavior of bond
percolation model but fail to reproduce the behavior for site percolation to which our ternary model reduces when $\varepsilon = 0$.
For the site model EMA procedures to obtain $g_{em}$ were proposed by 
Bernasconi and Wiesman in \cite{BW} and Yuge in \cite{Y}: in both
the usual effective medium procedure is varied in order to take into account the correlations 
between consecutive bonds naturally arising in this kind of systems. 
The fact that such correlations arise is easily understandable 
when considering a simple example:
let us take three neighboring sites $i$, $j$ and $k$ where  
$i$ and $k$ are two different nearest neighbors of $j$, and consider the bonds $ij$ and $jk$. 
If $j$ is, say, white \textit{none} of these bonds can have conductivity $g_2$ because both of them involve the white site $j$,
and their conductivities are not independent as assumed in the bond-based approach, and
the usual EMA technique has to be appropriately changed. Our approach here follows the lines of Ref. \cite{Y}.

Calculations start by
considering for any site of the lattice possessing a color index $\alpha=1,2,3$
(corresponding to white, red and black, respectively)
the mathematical expectation of the conductivity $\bar{g}_i$ of a bond starting from it:
\begin{eqnarray}
\bar{g}_1 &=& \phi_1g_1 + \phi_2g_3  \nonumber  \\
\bar{g}_2 &=& \phi_1g_3 + \phi_2g_2     \\
\bar{g}_3 &=& 0 \nonumber
\end{eqnarray}
These values appear in the system according to the 
the probabilities of their respective sites
\begin{equation}
P(\bar{g}) = \sum_{i=1}^{3} \phi_i \delta(\bar{g} - \bar{g}_i).
\end{equation} 
The effective conductivity is then obtained 
through the usual self-consistency condition \cite{K}
\begin{equation}\label{scc2}
\left\langle \frac{ g_{em} - \bar{g} }{(d-1)g_{em} +  \bar{g}} \right\rangle_P = 0.
\end{equation}
where $d$ is the dimension and $\left\langle \cdot \right\rangle_P$ is the average with respect 
to the distribution $P$ above.
If we now define a rescaled effective conductivity
\begin{equation}
f_{em}  =\frac{(d-1)}{g_0} g_{em}  
\end{equation} 
and introduce the arithmetic mean and the $\phi_k$-weighted average of the quantity 
$E_i = e^{-\beta U_i/2}$
\begin{equation}\label{averages}
\overline{E}  = \frac{1}{3}(1 + e^{-\beta \varepsilon/2}) 
\qquad \mbox{and} \qquad
\langle E \rangle = \phi_1 + \phi_2e^{-\beta \varepsilon/2}
\end{equation}
equation (\ref{scc2}) reduces to a quadratic equation for $f_{em}$,
\begin{equation}\label{fem}
 f_{em}^2 + 
b(\vec{\phi}, \varepsilon) \; f_{em}  +  
c(\vec{\phi}, \varepsilon) = 0
\end{equation}
with
\begin{eqnarray}
b(\vec{\phi}, \varepsilon) &=& \langle E \rangle \Big(3\overline{E}   - d \langle E \rangle\Big)   \nonumber  \\
c(\vec{\phi}, \varepsilon) &=&    \langle E \rangle^2 \Big(1 - d(1-\phi_3)\Big)e^{-\beta\varepsilon/2}. \nonumber
\end{eqnarray} 
The value of $D_{em}$ follows from the solution of this equation via 
\begin{eqnarray}\label{Deff}
D_{em}  &=&  \frac{a^2 g_{em}}{\langle n_i^0 \rangle } =
a^2 w_0 \cdot \frac{f_{em} }{(d-1)(\phi_1 + \phi_2e^{-\beta \varepsilon}) }  = \nonumber \\
&=& D_0 \cdot \tilde{D}_d(\vec{\phi}, \varepsilon)
\end{eqnarray} 
where $D_0 = a^2 w_0$ is the diffusivity of a lattice where all sites are white. 
The critical threshold at which $D_{em}$ vanishes 
can be obtained by setting $c(\vec{\phi}, \varepsilon)=0$
\begin{equation}\label{pc}
\bar{\phi}_3 = 1 - 1/d.
\end{equation} 

At variance with a classical binary Flory-Huggins situation, even in a random model with given $\phi_3$ it is hard to get 
an analytical estimate for $\phi_1$ and $\phi_2$ 
as functions of the known $\phi_3$ concentration. In a random case $\phi_2$ is the total perimeter density of black clusters in a 
site percolation (whose behavior is in principle known, but whose values have to be estimated numerically), 
and $\phi_1$ corresponds to the rest of the sites. This task gets even harder if the chains are present, 
and the numerical simulations show that the existence of the chains does matter. 
To overcome the problem, we simulate our polymer model first and
extract the numerical values of  $\phi_1$ and $\phi_2$ from these simulations. These numerical values are then used 
in the corresponding EMA calculations, whose predictions, in their turn, are compared with the results of simulations of diffusion.

\section{Pure percolation (binary) model}

It would be nice to know, how large is the typical error arising from disregarding the chain structure of black sites, and what is the role 
the chain length plays in the simplest case, namely in a percolation model with correlated black sites given by the chains.
In this model the red and the white sites are indistinguishable, they have the total number concentration $\phi_1=1-\phi_3$, and the result
of our previous consideration reduces to the original Yuge's result for site percolation. This is exactly the situation discussed in the 
present section. 

\begin{figure}[h]
\begin{minipage}{4.0cm}
\begin{tabular}{c}
\hspace{-0.7cm}
\includegraphics[height=4.5cm]{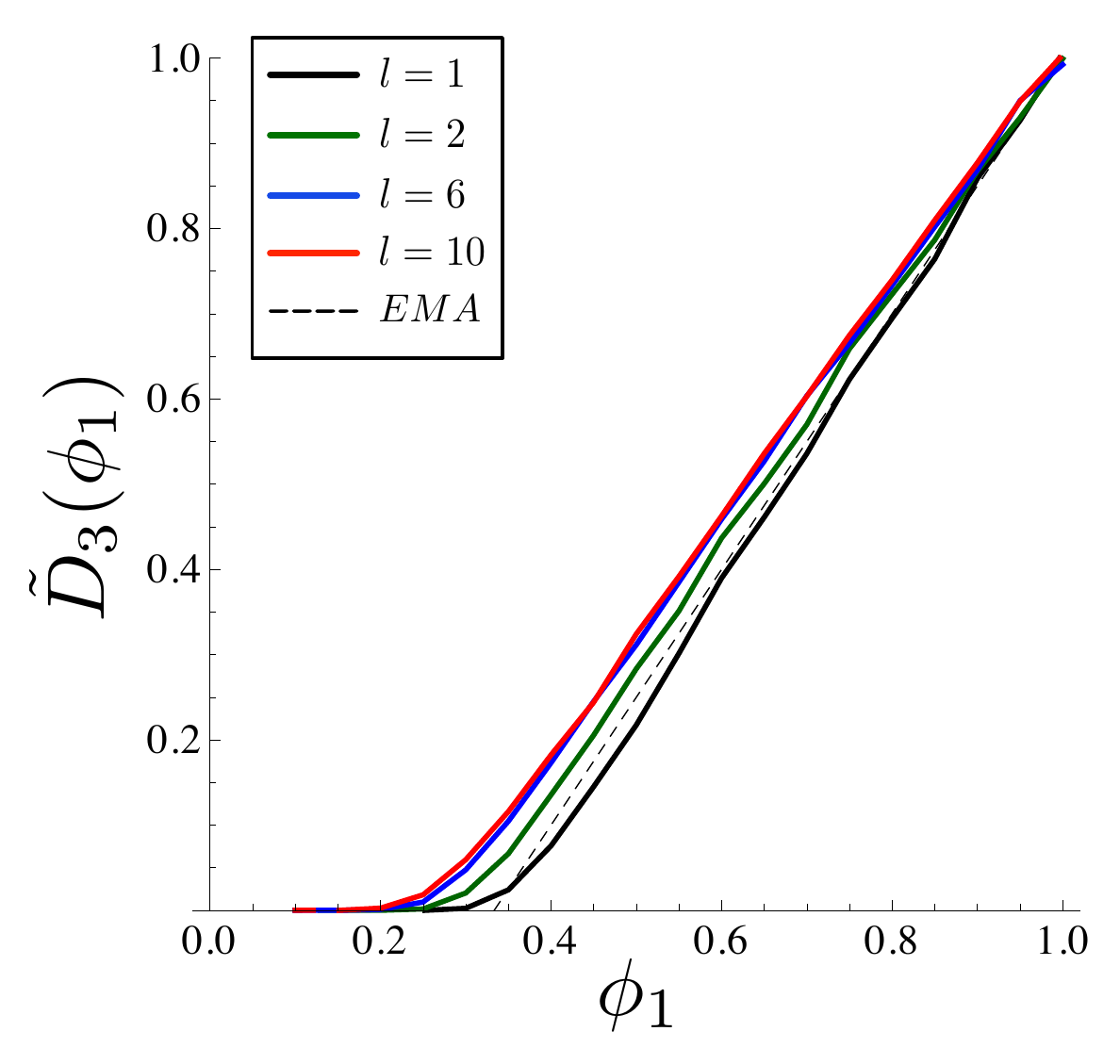}\\
(a)
\end{tabular}
\end{minipage}
\hfill
\begin{minipage}{4.0cm}
\begin{tabular}{c}
\hspace{-0.7cm}
\includegraphics[width=4.5cm]{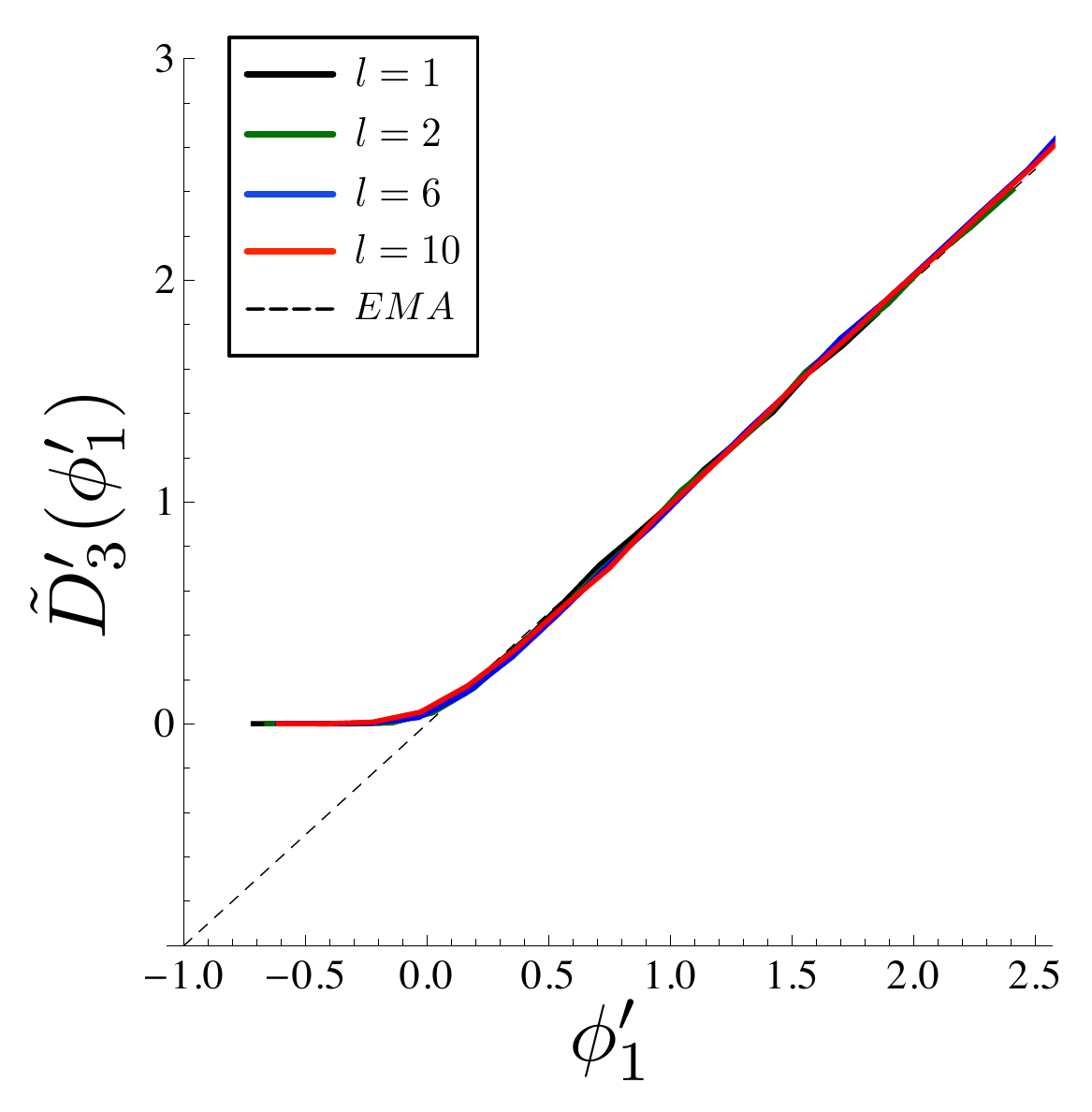}\\
(b)
\end{tabular}
\end{minipage}
\caption{(color online) (a) Normalized diffusivities in the purely percolation case vs.  the concentration $\phi_1$ of white sites; 
(b) Rescaled normalized diffusivities vs. rescaled white concentration $\phi_1^{'}$. 
The dotted line represents in both figures the effective medium approximation.}
\label{fig:A}
\end{figure}

Thus we consider a pure percolation situation in which the only interactions are the excluded volume ones and 
our lattice consists of only black and white sites, red ones are absent. The results of simulations
for the systems of chains of different lengths are shown in Fig. \ref{fig:A}(a). 
The figure representing the dependence of the diffusion coefficient on the concentration of 
sites occupied by segments of the chain shows this for the chain lengths from $l=1$
(usual Bernoulli site percolation problem) to $l=10$. The simulations were performed also for 
longer chains, but for $l$ larger than $10$ the corresponding graphs are indistinguishable 
from that for $l=10$ within the statistical accuracy. Thus, a result for $l=100$, 
(not shown) is indistinguishable from the one for $l=10$ on the scales of Fig \ref{fig:A}(a). 

Details about simulations are readily given: 
Simple random walks of $l$ steps 
are let run independently in a lattice of $400^3$ sites with 
periodic boundary conditions. 
This operation is stopped when the total segment concentration of segments (sites visited at least once) is 
within $0.01$ from the desired value of $\phi_3$. 
Once the environment is created, $10^6$ random walks of $10^3$ to $10^4$ steps, 
depending on the speed of homogenization of the system, are launched from a 
free site chosen at random in a cube of $50^3$ sites placed in the center of the medium. 
With this choice, the probability for a diffusing particle to reach 
the borders of the lattice is extremely low and doesn't spoil the statistics. 
The algorithm used is the Monte Carlo Blind Ant one. 
The whole procedure is then repeated for $10$ different lattice realizations and averages are taken.
We have observed a normal diffusion process $\langle r^2(t)\rangle \propto t$  
from which the proportionality constant $\tilde{D}_3$ has been extracted and reported in Fig. \ref{fig:A}(a). 
The homogenization of $\langle r^2(t)\rangle$ slows down in the proximity of the critical point. For this reason 
$10^4$ time steps become insufficient and the
diffusivity is systematically  overestimated. Our attention however is focused 
on a range of values of $\phi_3$ which are above the percolation threshold.

The curves do not differ drastically, but definitely show different percolation thresholds
$\phi_1^{c}(l)$ depending on $l$. 
For the Bernoulli case the total behavior of diffusivity is reproduced sufficiently well by EMA for 
$\phi_3$ close to unity but departures from the EMA line for concentrations close to a critical one.
For longer chains the critical concentration gets lower, and the diffusion coefficient at given $\phi_1$ gets larger than
for the Bernoulli case.
Although different, the curves however show a large amount of universality 
which is unveiled when rescaling the concentration and diffusivity according to 
\begin{equation}  
\phi_1^{'} = \frac{\phi_1}{\phi_1^{c}} - 1 
\qquad \mathrm{and} \qquad
\tilde{D}_3^{'} =  \tilde{D}_3\frac{(1-\phi_1^{c})}{\phi_1^{c}},
\end{equation}
so that the critical concentration is mapped onto the point $\phi_1^{'} = 0$, see \ref{fig:A}(b). 
In this case all the curves fall onto the same master curve, and the mean-field result, rescaled 
accordingly, gives a straight line (of slope 1) which reproduces the results of simulations astonishingly well up to
the critical domain. This high degree of universality shows that the correlations introduced by 
the existence of the chain are not of high importance and can be fully accounted for by rescaling the results of EMA
according to the equations above. The corresponding critical concentration has however to be obtained numerically. Alternatively, it
can be extrapolated from the slope of diffusion coefficient for concentrations close to unity.

\section{Results for ternary model}

In this section we discuss results for the normalized 
effective diffusivity $\tilde{D}_3(\vec{\phi}, \varepsilon)$ 
and concentrate on the role of interaction energy $\varepsilon$ between the diffusing particles and the polymer matrix.
All the figures refer to the three-dimensional case.
The reduced interaction energy $\beta \varepsilon = \bar{\varepsilon}$ 
is chosen to span in the interval $[-5, 5]$ according to the following 
reasoning: typical absolute values of $\varepsilon_{XX}/K_B$, the coupling strength 
of a Lennard-Jones potential describing the interaction between two atoms 
of the same kind X, can be roughly enclosed in the interval corresponding to temperatures $[0, 500 K]$. 
In order to consider the interaction between two different atoms X and Y 
the Lorentz-Berthelot mixing rule is used to obtain
$\varepsilon_{XY} = \sqrt{\varepsilon_{XX}\varepsilon_{YY}}$  which,
being an average, belongs to the same interval. 
Using $\varepsilon$ in place of $\varepsilon_{XY}$, 
considering both positive and negative values and 
taking the temperature not too far from the ambient one,  
it is straightforward to see that the choice $\bar{\varepsilon} \in [-5, 5]$   
is a reasonable one. For the discussion of the Arrhenius-like or non-Arrhenius temperature dependencies
in Sec.~\ref{Arrhe} 
broader bounds are used, $\bar{\varepsilon} \in [-10, 10]$

\subsection{Effective diffusivity vs interaction energy}

\begin{figure}[h!]
\hspace{-1.0cm}
\begin{minipage}{3.0cm}
\begin{tabular}{c}
\includegraphics[height=3.0cm]{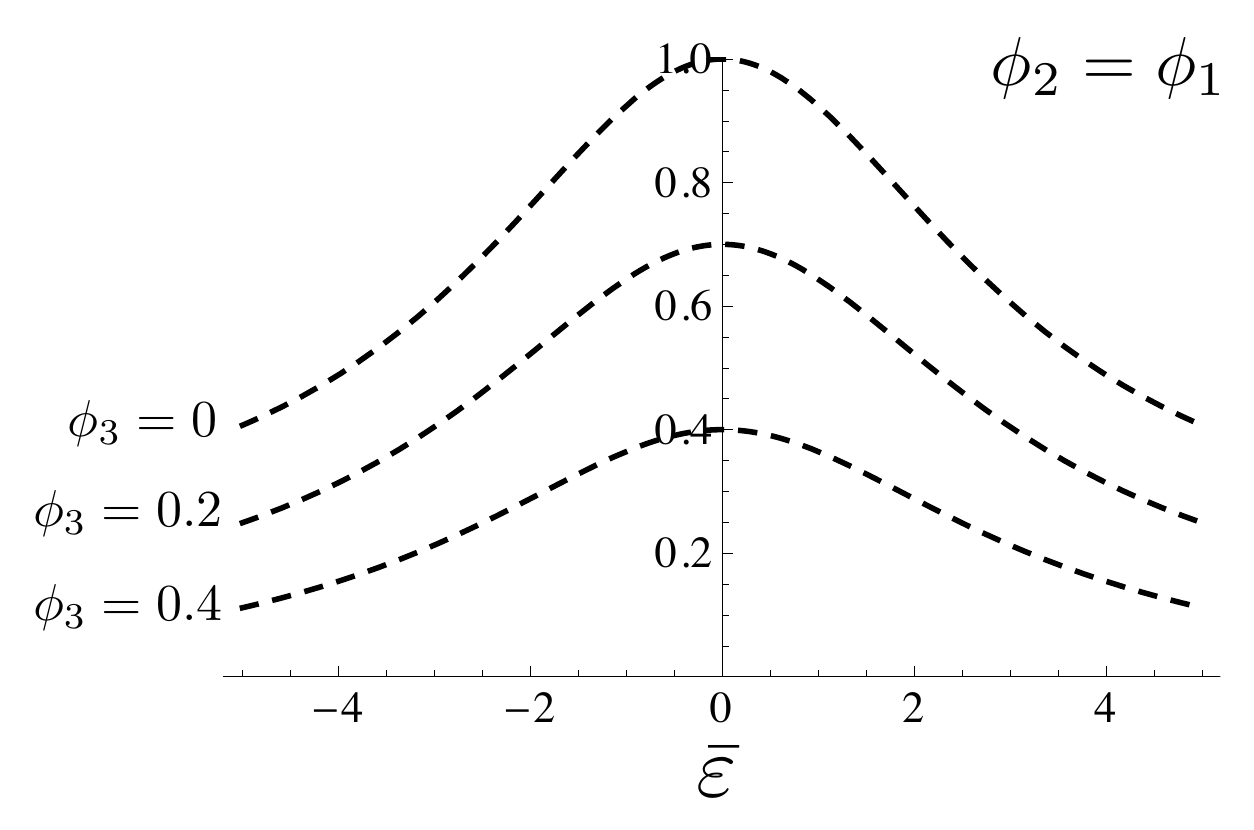} \\
(a)
\end{tabular}
\end{minipage}
\hfill
\begin{minipage}{3.0cm}
\begin{tabular}{c}
\hspace{-1.5cm}
\includegraphics[height=3.0cm]{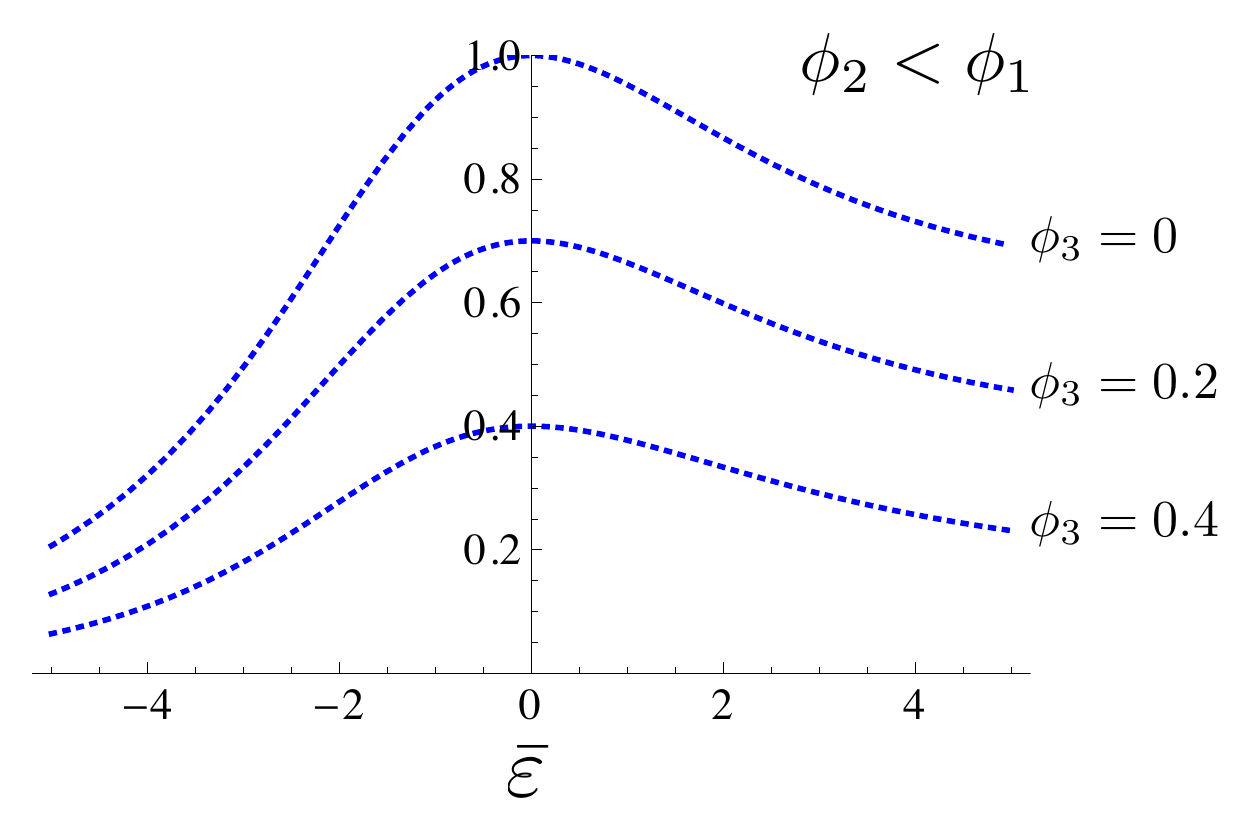} \\
(b)
\end{tabular}
\end{minipage}
\\
\hspace{-1.0cm}
\begin{minipage}{3.0cm}
\begin{tabular}{c}
\includegraphics[height=3.0cm]{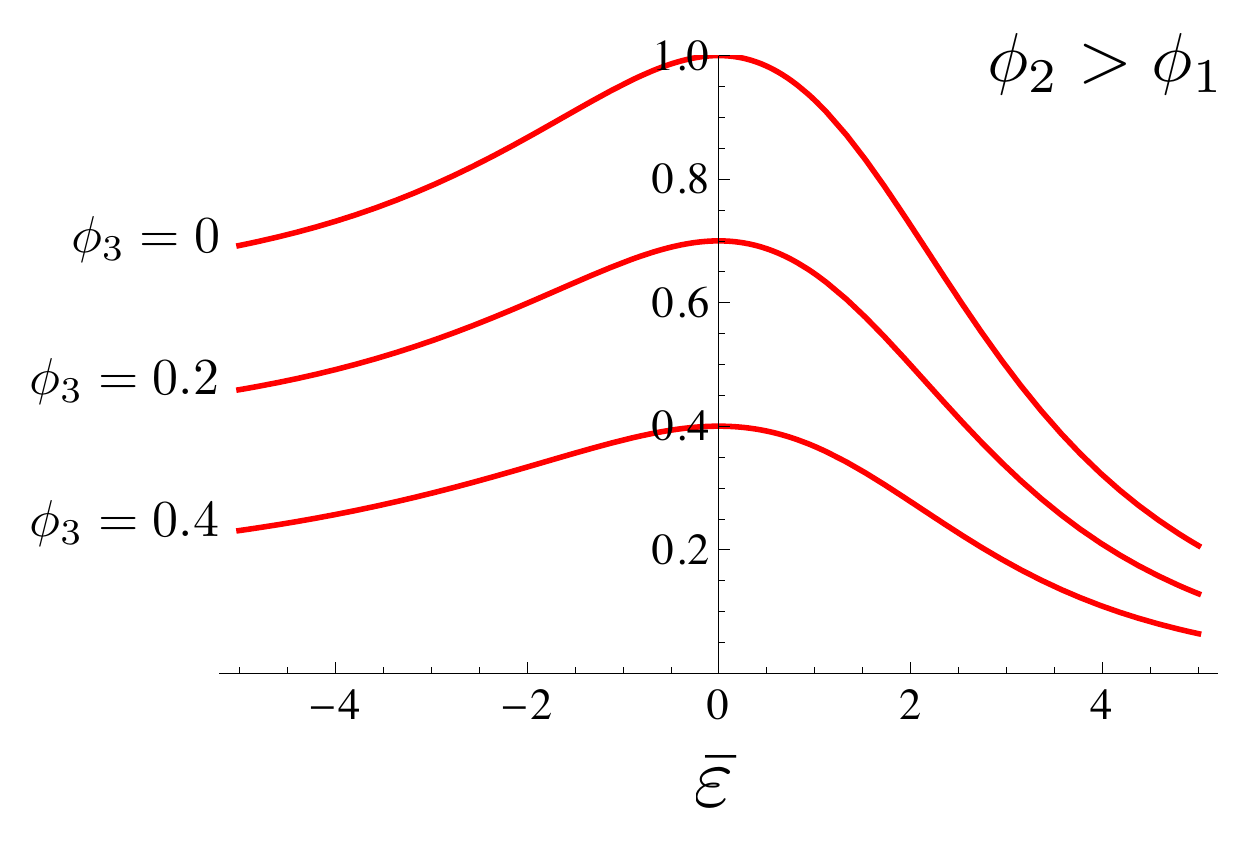}\\
(c)
\end{tabular}
\end{minipage}
\hfill
\begin{minipage}{3.0cm}
\begin{tabular}{c}
\hspace{-1.9cm}
\includegraphics[height=3.0cm]{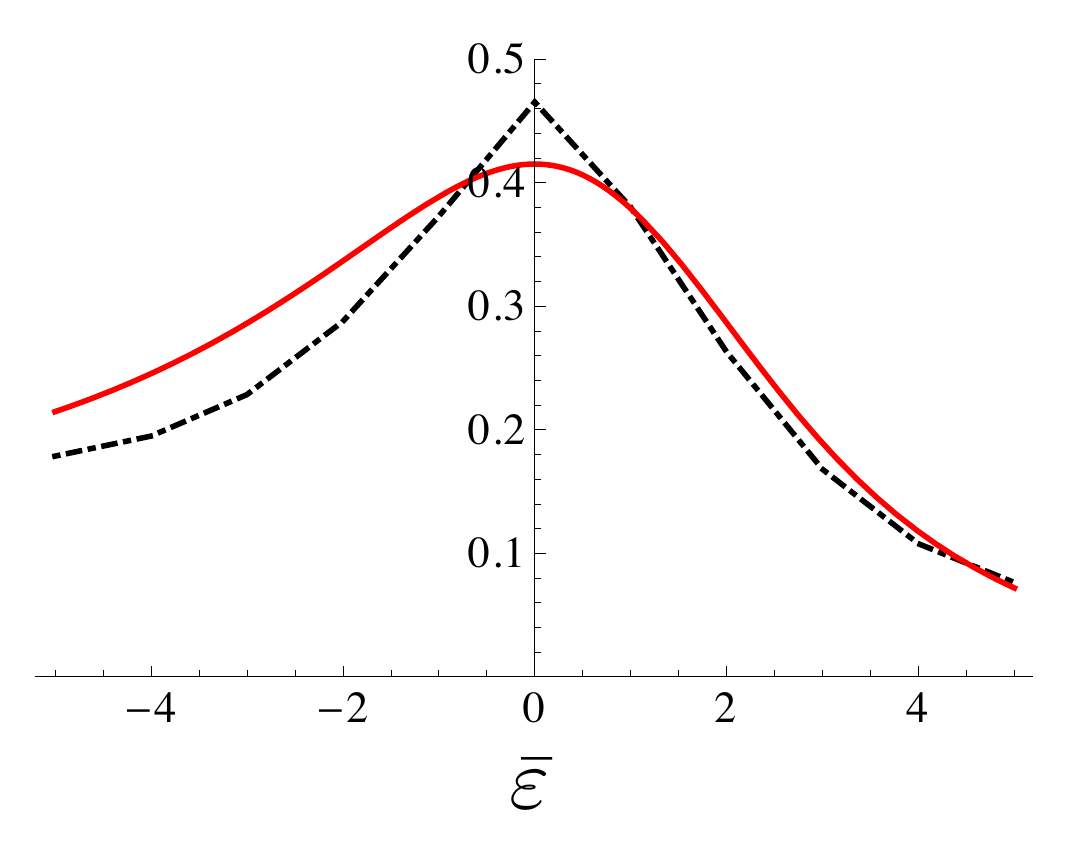}\\
(d)
\end{tabular}
\end{minipage}
\caption{(color online) Mean field normalized effective diffusivity $\tilde{D}_3$ vs $\bar{\varepsilon}$  
in the case $\phi_3=0; 0.2; 0.4$ and: (a) $\gamma=1/2$;  (b) $\gamma=3/4$;  (c) $\gamma=1/4$. 
(d) Comparison between theory and simulations (black dash-dotted line) in the case $\phi_3=0.39$, $\phi_2=0.43 \; (\gamma = 0.295)$.}
\label{fig:D}
\end{figure}

Let us first discuss general features of the dependence of the diffusion coefficient on 
number concentrations and on interaction energy $\varepsilon$.
The EMA results for $\tilde{D}_3(\vec{\phi}, \varepsilon)$ 
for the three different cases corresponding to different relations between $\phi_1$ and $\phi_2$ for 
$\phi_3$ fixed are shown in Fig. \ref{fig:D}. 
These plots show the behavior
for the attractive and repulsive interaction and the way the diffusivity approaches zero when the black sites concentration approaches 
its critical value $\bar{\phi}_3 =2/3$ (see Eq.(\ref{pc})). 
At this value in fact, particles remain confined in finite 
subregions of the system, due to the overwhelming predominance 
of polymer segments.

Plots are given for three different sets of the $\phi_k$ values 
in order to consider symmetrically the situations in which
red sites are in minority, equally probable or predominant with respect to the white ones, 
at given $\phi_3$. 
For this purpose we introduce a real parameter $\gamma \in [0,1]$ 
defining the number concentrations of white and red sites as
\begin{eqnarray}
\phi_1 &=& \gamma(1-\phi_3) \\
\phi_2 &=&(1-\gamma)(1-\phi_3). 
\end{eqnarray}
Graphs are then taken for three different values of $\gamma$ (color online): 
$\gamma=3/4$ (blue dotted lines, $\phi_2<\phi_1$), 
$\gamma=1/2$ (black dashed  lines, $\phi_2=\phi_1$) 
and $\gamma=1/4$ (red solid lines, $\phi_2>\phi_1$).
This imbalance will deeply influence the behavior of the effective diffusivity 
when $\varepsilon$ crosses the zero value.

In the symmetric case $\phi_2=\phi_1$, 
$\tilde{D}_3$ is invariant under the change of the sign of interaction energy $\bar{\varepsilon} \rightarrow - \bar{\varepsilon}$.
On the contrary, when the white-red balance is broken, the 
effective diffusivity decreases or increases 
depending on the sign of the energy parameter and on 
the value of $\gamma$. 
Let us consider the situation
in which $\phi_2>\phi_1$ (e.g. $\gamma=1/4$,  Fig. \ref{fig:D}(c)) 
and restrict our attention on 
the attractive $\bar{\varepsilon}<0$ region;
with this choice, we increase the number 
of the red-red $g_2$ bonds (showing larger conductivity) with respect to the number of the white-white $g_0$ ones 
which have the lowest conductivity. This results in a global increasing of the effective diffusion constant.
If we now invert the sign of $\varepsilon$, i.e. consider the repulsive interaction, the $g_2$ bonds will 
still be the most numerous, but now have lowest conductivity value, 
decreasing in this way the whole diffusivity of the system. 
The opposite happens if we consider $\phi_2<\phi_1$; the corresponding graph in Fig.\ref{fig:D}(b) is a 
mirror image of the one in Fig. \ref{fig:D}(c).
The comparison between the mean field calculations and the
Monte Carlo simulations performed in the original ternary 
lattice corresponds to the chain length $l=100$ is quite satisfactory (fig.\ref{fig:D}(d)).
Once the desired polymer concentration $\phi_3 \simeq 0.4$ was reached 
and the polymer matrix is set up, the energy value $\varepsilon$ is assigned to
all the nearest neighbors of the segments and their concentration $\phi_2$ is measured. 
All results are averaged over 10 realizations of the polymeric matrix.
In each of them random walks of $10^4$ steps were performed as described above.
The number of realizations per energy landscape is $10^6$ times. 
The numerical values of number concentrations are $\phi_3=0.39$ and $\phi_2=0.43$.  
The numerical  result is then plotted together with the mean field calculations 
in which the same values are used (figure \ref{fig:D}(d)).
We note that the value of the polymer concentration is close to the critical domain in Fig. \ref{fig:A}(a) corresponding
to $\varepsilon=0$, so that the total accuracy of EMA is not too high in this domain. However, the EMA-results reproduce the dependence 
qualitatively well, and, moreover, the accuracy of EMA improves for higher interaction strengths.

\subsection{Arrhenius vs. non-Arrhenius behavior}
\label{Arrhe}
\begin{figure}[h]
\begin{minipage}{4.0cm}
\begin{tabular}{c}
\hspace{-1.5cm}
\includegraphics[height=4.0cm]{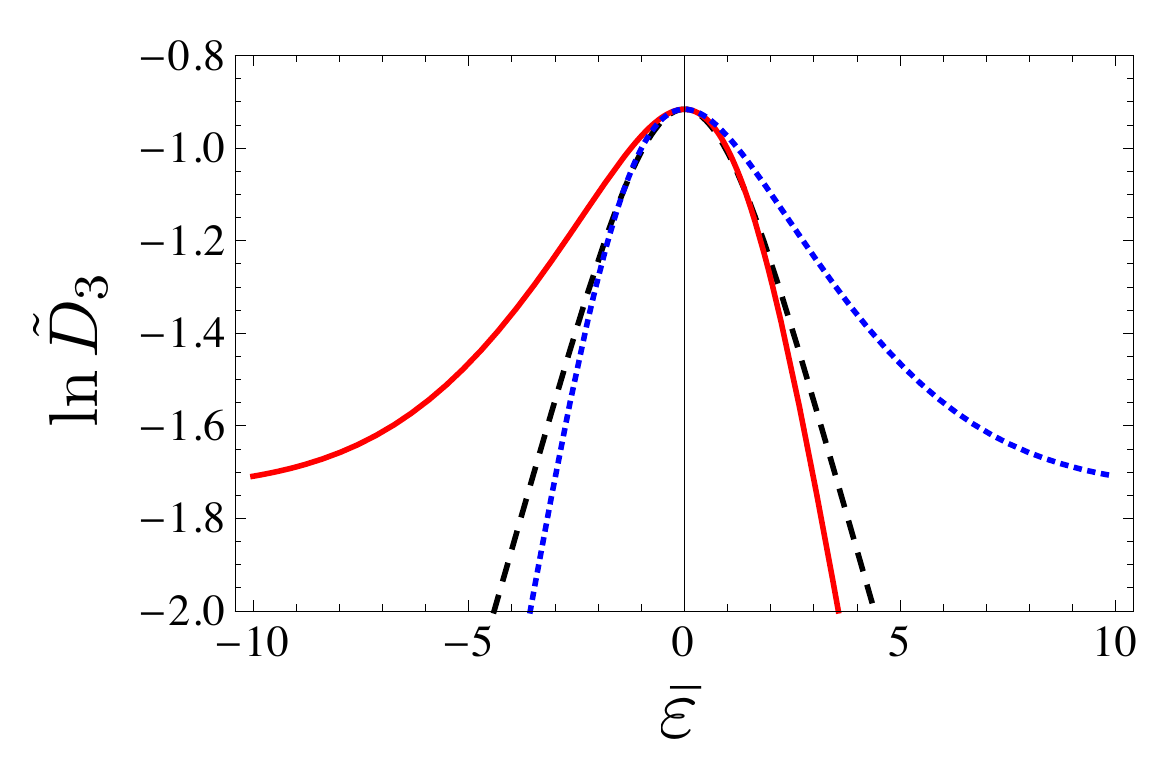} \\
(a)
\end{tabular}
\end{minipage}
\vfill
\begin{minipage}{4.0cm}
\begin{tabular}{c}
\hspace{-1.4cm}
\includegraphics[height=4.0cm]{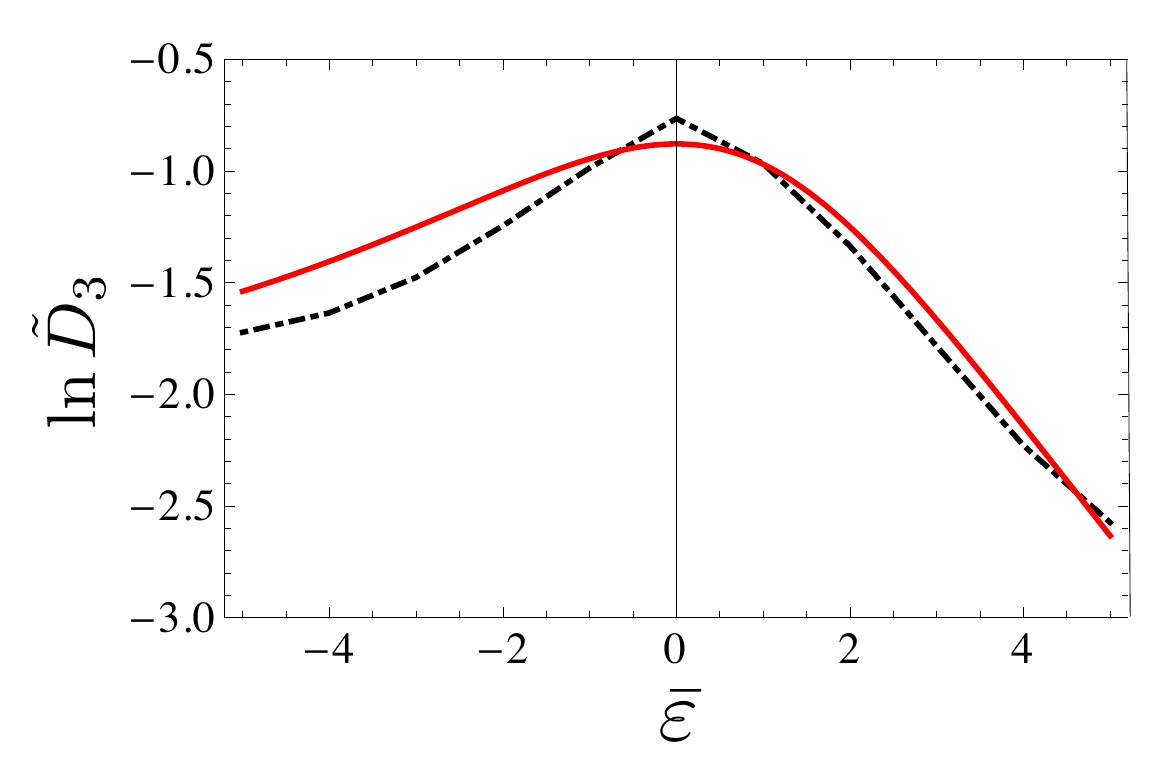}\\
(b)
\end{tabular}
\end{minipage}
\caption{(color online) (a): Arrhenius plots of the different mean field normalized 
effective diffusivities vs. $\bar{\varepsilon}$ at $\phi_3=0.4$.
(b):Comparison between theory and simulation (black dash-dotted line) 
Arrhenius plots at $\phi_3=0.39$, $\phi_2=0.43$.}
\label{fig:arr}
\end{figure}

A non-trivial aspect of the dependence of diffusivity on the interaction strength is revealed by the 
Arrhenius plots shown in figure \ref{fig:arr}(a) where
the logarithm of $\tilde{D}_3$ is plotted as a function of $\bar{\varepsilon} = \varepsilon/K_BT$ 
in the wider interval $[-10,10]$, to investigate the role played by activation in the diffusion process;
the segment concentration is set here to $\phi_3=0.4$. The three curves in Fig. \ref{fig:arr}(a) correspond to the values of $\gamma = 1/4$,
$\gamma = 1/2$ and $\gamma = 3/4$. As in the previous figures, the curve for $\gamma = 1/2$ represents an even function of $\bar{\varepsilon}$,
and the curves for $\gamma = 1/4$ and $\gamma = 3/4$ are mirror images of each other.
For $\bar{\varepsilon}$ close to zero, 
the activation process is not relevant, the curves
fall together and reproduce the diffusion constant in the black-and-white lattice of section 4. 
When moving away from the $\bar{\varepsilon}=0$ value,  
the activation acquires importance. For $\gamma = 1/2$ this behavior becomes Arrhenius-like 
and the curve shows a linear decay for both signs of $\bar{\varepsilon}$ provided the interaction is strong enough. For  asymmetric cases 
$\gamma \neq 1/2$ the Arrhenius behavior is seen only for interaction energy of the corresponding sign (attractive interaction for 
$\gamma < 1/2$ and repulsive interaction for $\gamma > 1/2$).
For the opposite sign of interaction, at low temperatures, or high absolute values of $\varepsilon$, 
the lines become horizontal, quitting the Arrhenius regime. 
This non-Arrhenius behavior can be explained as follows.
Let us focus our attention again on the red (solid) line in the negative $\bar{\varepsilon}$ half-plane.
Under segment concentration $\phi_3=0.4$ 
the black infinite cluster exists but is not dense enough to prevent the existence of infinite white or red ones.
The concentration of red sites is $\phi_2 = 0.45$ $(\gamma=1/4)$ and thus lays
above the percolation threshold for a cubic lattice. This means that 
red sites form an infinite cluster crossing the whole
system, and once a particle finds it, it can travel on it through the whole system 
rather than escape from it by activation. As a consequence 
diffusivity saturates and the system never freezes. 
In the repulsive region, the same behavior 
is shown by the blue (dotted) line, indicating the existence of a white infinite cluster. The  black (dashed) line, 
the one for symmetric situation $\phi_3=0.4$, $\phi_2=\phi_1=0.3$, doesn't show any 
saturation. This suggests that in such a case white and red concentrations are 
below the percolation threshold, and the activation processes are necessary to traverse the system.

Figure \ref{fig:arr}(b) shows the comparison between theory and simulation Arrhenius plots 
in the original interval $\bar{\varepsilon} \in [-5,5]$.

On the total the following regimes of behavior 
can be qualitatively distinguished:

(1) If the concentration of black sites is so high that percolation on 
red and white sites is not possible, the diffusion coefficient vanishes.

In the case when percolation over the red-and-white domains is possible, 
the diffusion coefficient is nonzero, and its behavior as a function of temperature  
depends on the percolation properties of red and white clusters, 
and on the sign of interaction energy.

If the interaction is repulsive, two regimes appear:

(2) If white sites percolate, the diffusion over the white cluster is always 
possible and does not need activation. The temperature dependence saturates.

(3) If white clusters do not percolate, the diffusion is only possible over red sites, 
and involves an activation process; its temperature dependence shows the Arrhenius behavior.

In the case of attractive interaction the roles of white and red sites interchange, 
and percolation over red sites is what determines the 
temperature dependence of the diffusion coefficient:

(4) If red sites do percolate, the diffusion over the red cluster is 
possible and does not need activation. The temperature dependence saturates.

(5) If red clusters do not percolate, the diffusion has to go via white sites, 
and therefore involves an activation process; 
its temperature dependence shows the Arrhenius behavior.

\begin{figure}[h]
\begin{center}
\includegraphics[height=4.0cm]{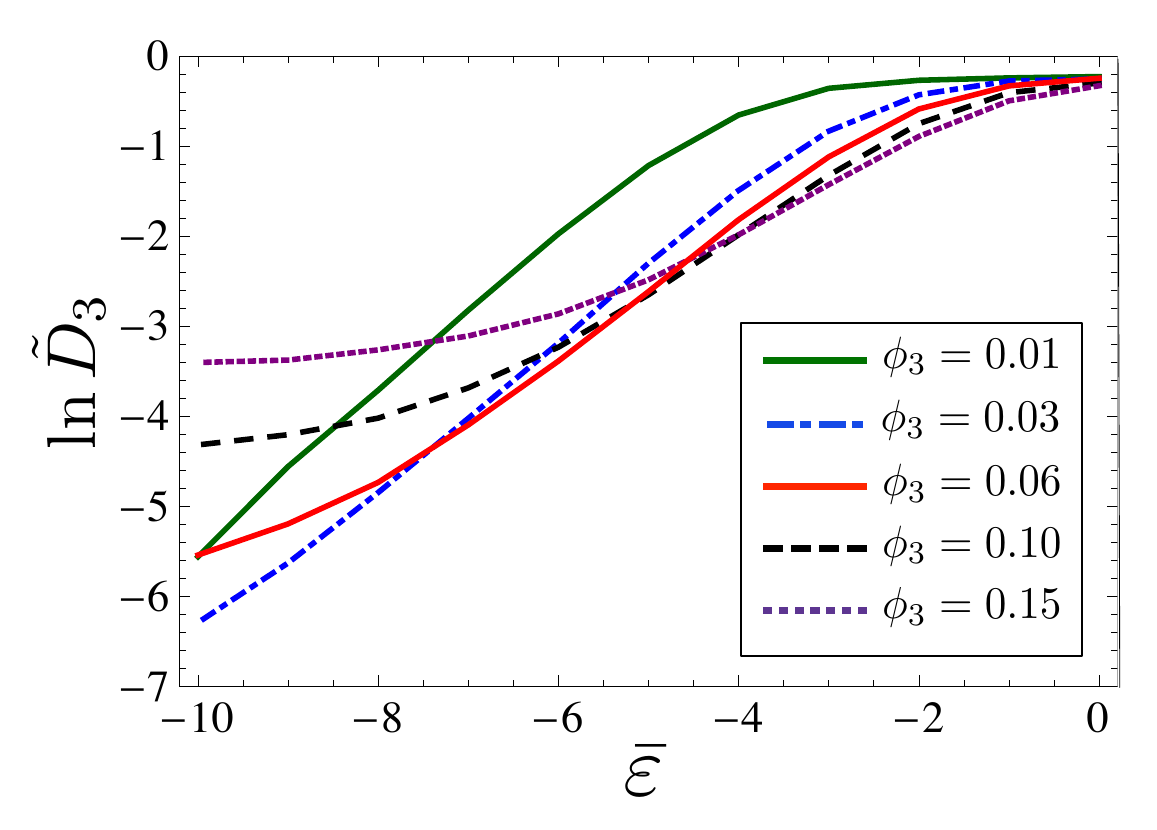} 
\caption{(color online) Simulation graph: Arrhenius plots of the normalized diffusivity in the attractive case 
vs. $\bar{\varepsilon}$ for different segment concentrations.}
\label{fig:sim}
\end{center}
\end{figure}

These features, predicted by EMA, have also been found in simulations of a genuine ternary lattice 
in which red clusters run clung on the black chains by construction. 
Figure \ref{fig:sim} shows the behavior of Arrhenius plots for low polymer concentrations in the case of attractive interaction.
It shows the logarithm of the normalized effective diffusivity for different values of $\phi_3$. 
For $\phi_3<0.06$, polymers remain sparse and isolated, 
their red perimeter sites don't percolate, no infinite red cluster 
exists and the system is in an Arrhenius regime (5).
When the number of chains is increased, the transition from the Arrhenius to the saturation behavior (4)
is observed at the critical value $\phi_3=0.06$, revealing the emergence of an infinite red cluster. 
This critical value is far below the usual percolation threshold of a cubic
lattice due to the fact that red sites are arranged in connected groups on the perimeters of black chains.
This number can not be predicted by simple EMA and can be translated into an estimate 
of the percolation threshold of perimeter sites of chains. 

\section{Conclusions}

We have considered diffusion of small molecules in a solid polymeric medium taking into account the interaction 
between polymers and diffusing particles which can be both attractive or repulsive.  
The diffusivity has been analyzed from different perspectives both analytically, 
using a modified effective medium approximation, and numerically 
by performing direct Monte Carlo simulations. While the diffusivity is only slightly affected by the chain's length,
its temperature dependence crucially depends on the kind of interaction. This behavior depends on the sign of the interaction energy
and is related to the existence of a percolating cluster of interaction sites surrounding polymer segments and/or a percolating cluster of
voids on which particles are free to travel without activation.


\begin{center}
\textbf{ACKNOWLEDGMENTS}
\end{center}
The work was supported by BMU within the  
project "Zuverl\"assigkeit von PV Modulen II".

\end{document}